\documentclass[prd,showpacs,amsmath,showkeys,twocolumn,floatfix]{revtex4} 
\usepackage[]{epsfig,dcolumn}
\usepackage[hang,normalsize,bf]{subfigure} 
\usepackage{graphicx}
\usepackage{bm}

\def\L{{\bf L}}

\def\lsim{\mathrel{\rlap{\lower4pt\hbox{\hskip1pt$\sim$}}
    \raise1pt\hbox{$<$}}}
\def\gsim{\mathrel{\rlap{\lower4pt\hbox{\hskip1pt$\sim$}}
    \raise1pt\hbox{$>$}}}

\begin{document}

\title{
Diffractive photoproduction of opposite-charge
pseudoscalar meson pairs at high energies }

\author{Antoni  Szczurek$^{1,3,4}$ and Adam P. Szczepaniak$^{1,2}$ }

\affiliation{$^{1}$ Nuclear Theory Center, Indiana University, Bloomington, Indiana 47405, USA\\
$^{2}$  Physics Department,  Indiana University, \\ 
Bloomington, Indiana 47405, USA\\
$^{3}$ Institute of Nuclear Physics\\
PL-31-342 Cracow, Poland\\ 
$^{4}$ University of Rzesz\'ow\\
PL-35-959 Rzesz\'ow, Poland }

\date{\today}

\begin{abstract}
We calculate the cross section for diffractive photoproduction
of opposite-charge pseudoscalar meson pairs $M^+M^- =$  
$\pi^+ \pi^-$, $K^+ K^-$, $D^+ D^-$ and $B^+ B^-$
in a broad range of center-of-mass energies 
relevant for GlueX/Hall D, FOCUS, COMPASS and HERA experiments.
In the case of $\pi^+\pi^-$ production we find that the 
 interference of the $\rho^0$ resonance and the two-pion continuum leads to a considerable deformation of the shape of $\rho^0$ in agreement with the data from the ZEUS collaboration.
%The effect is almost independent of the incident energy.
We also discuss  the spectral shape of the $\rho^0$ as a function of
the momentum transfer and  the contribution
of higher partial waves to the $\pi^+\pi^-$ mass spectrum. 
%We study the energy-dependent forward-backward angular asymmetry
%of $M^+$ and $M^-$ in the recoil center of mass (RCM) distributions.
We predict a sizeable energy-dependent forward-backward asymmetry in
the Gottfried-Jackson frame.
For the heavy meson production we find that the cross section for
diffractive production increases much slower than the one for
open charm or bottom production.
We discuss lower and upper limits for the cross sections for
diffractive production of $D^+ D^-$ and $B^+ B^-$ pairs,
which we find can be as large as 10\% of the open flavor production.
\end{abstract}

\pacs{13.60.Le, 11.80.Et, 12.40.Nn}

\maketitle

%---------------------
\section{Introduction}
%---------------------

The mechanism of diffractive photoproduction of charged pion pairs
has been proposed long ago
\cite{Drell60,Soding65,Krass67,KU69,Pumplin70,KQ71}.
At that time there, however, no good data was available for constraining details of the model. 
% and also the necessary ingredients of the model
%calculations were rather poorly known.
In recent years a body of data from $\pi N$ and $K N$ scattering
has been collected (see for instance \cite{PDG}).
In particular  the ZEUS and H1 collaborations at HERA
have measured diffractive production of the $\rho^0$ meson.
Most of the theoretical effort, however, concentrated on the description of the total cross section in different pQCD inspired approaches and not 
 on the description of the two-pion continuum invariant mass distribution.

In real photoproduction the ZEUS collaboration observed
a strong 
%reversed 
 asymmetry  in the two-pion invariant mass
 spectrum around the peak position \cite{ZEUS_Mpipi} of the $\rho^0$. 
In electroptoduction the asymmetry seems to decrease with increasing photon virtuality \cite{ZEUS_Q2}, $Q^2$ and momentum transfer
 $t'=|t-t_{min}|$~\cite{ZEUS_t}. One would expect, whatever the  
  non-resonant mechanism might be, that it will also be present at small $Q^2$ and small $t'$ and thus needed to isolate the $\rho^0$ production cross section. 
  
  The GlueX (Hall D) project at the Jefferson Lab will 
 study mesonic resonances focusing on gluonic excitations. 
   The diffractive production of
  charged meson continuum  may produce  large  backgrounds to the   channels of interest.  Recently exotic
$J^{PC}$ = 1$^{-+}$ candidates have been reported by the E852 \cite{E852}
and the Crystall Barrel \cite{CB} collaborations. Candidates for the exotic
states are rather broad \cite{E852,CB} and may indeed have
 a large component originating from production of
 the meson continuum production~\cite{AS1,AS2}.  
 
 Recently the FOCUS collaboration at Fermilab found a new state
 in the K$^+$ K$^-$ final state at 1.75 GeV \cite{FOCUS_KK} wich 
  may also require a good understanding of the $K{\bar K}$, continuum production.  
%3$\pi$, 2K, $\eta \pi$, $\eta' \pi$, etc., continua will be necessary
%to confirm the existence of the new states.

In the present paper we study the $\pi^+ \pi^-$ and $K^+ K^-$ channels.
 In this case model ingredients
 are well constrained by the $\pi N$ and $K N$ data. The $\pi^+ \pi^-$  invariant mass distribution
and possible $\pi^+ - \pi^-$ polar angle asymmetries are
presently being analyzed by the COMPASS collaboration \cite{COMPASS}. 
We also compare the results of our calculation with
the experimental data at higher energies from HERA and medium energy 
  relevant for the future GlueX/Hall D experiment 
 at TJNAF. 

 The inclusive production of heavy charmed mesons in electro- 
 or pho-production off proton is routinely used to study the gluon
 distribution in the nucleon. The standard QCD approach is based
 on the production of heavy quark -- heavy antiquark pairs
  at the parton level from photon-gluon fusion followed by a fragmentation 
   to heavy flavored hadrons. The formalism which we present 
  for diffractive production of pairs of light charged mesons should be also valid for the production of pairs of heavy mesons, $D^+ D^-$ or even $B^+ B^-$. This may be interesting in the context of a deficit in the open $b \bar b$ production in photon-proton \cite{gp_bbbar} and photon-photon \cite{gg_bbbar} collisions.

Recently the FOCUS collaboration has analyzed
the azimuthal correlations between $D \bar D$ mesons \cite{FOCUS_DD}.
It was pointed out very recently \cite{LS04} that heavy meson
correlations are very useful to study
unintegrated gluon distributions in the nucleon.
To the best of our knowledge, the contribution of the  diffractive mechanism, discussed here, has not been estimated in this context. 
It is also interesting to investigate how large the diffractive production of $B^+ B^-$ pairs might be, compared to the standard pQCD mechanism
of $b \bar b$ production discussed above. 
In photoproduction, one would naively expect a relative enhancement
of the ratio of diffractive $B^+ B^-$ (charge 1) to the standard
$b \bar b$ (charge 1/3) production as compared to the ratio of
the $D^+ D^-$ (charge 1) to the $c \bar c$ (charge 2/3) production
by a factor of 4 in the cross section.
A better understanding requires, however, more detailed insight
into the dynamics of the process.
In the present paper we shall present estimates of
such contributions. In particular, we discuss several aspects of
the opposite-charge pseudoscalar continuum in photoproduction.

We note that a similar model has been recently applied to describe $\pi\pi$ and $K{\bar K}$ photoproduction  in Ref.\cite{JKLSW1998}.
%We discuss a very simple (one free parameter) model which is powerful
%to describe the existing experimental data. We make several predictions
%for the present and forthcoming experiments.  

%----------------------------------------------
\section{Model of the continuum}
%----------------------------------------------

\begin{figure}[htb] % Figure 1
%\begin{center}
\includegraphics[width=3.0cm]{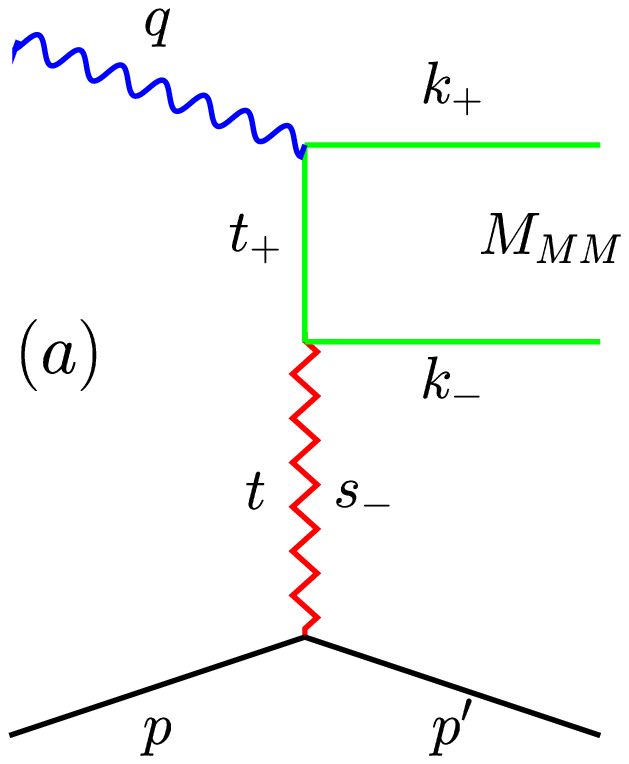}
\includegraphics[width=3.0cm]{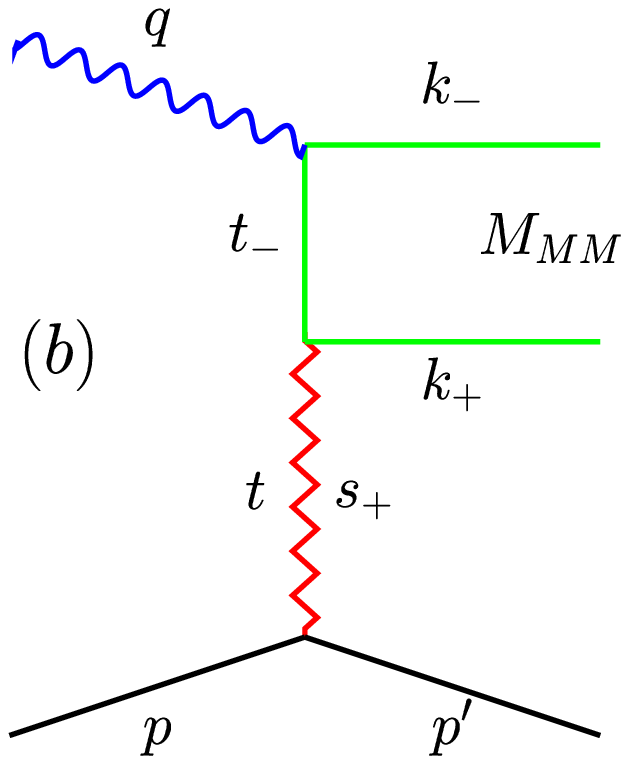}
%\\[30pt]
\includegraphics[width=3.0cm]{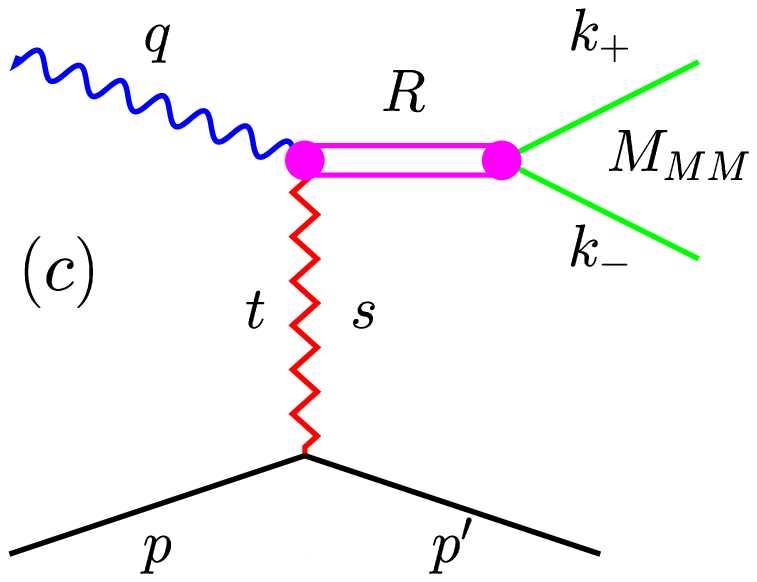}
\caption{
Diffractive photoproduction of opposite-charge
meson pairs. The wavy line corresponds to the photon, the zigzag
line describes the pomeron and subleading reggeon exchanges,
the lower lines are incoming and outgoing nucleons.
\label{fig_diagrams} }
%\end{center}
\end{figure}

The dominant mechanisms of diffractive production of opposite-charge
meson pairs are shown in Fig.\ref{fig_diagrams}.
%------------------
The continuum production shown by the diagrams (a) and (b) in
Fig.\ref{fig_diagrams} is often a background  to direct resonance
($\rho^0$, $\phi$, $f_2$, etc.) production shown in 
diagram (c). We shall refer to (a)-(b) and (c) as the continuum and
resonance contributions, respectively. The zigzag line represents
the pomeron and subleading reggeon exchanges.
The cross section for diffractive photoproduction of the opposite
charge meson pairs, 
%which includes besides the three-body phase space
can be written as
\begin{widetext}
\begin{equation}
d \sigma  =  (2 \pi)^4 \delta^4(q + p - p_{+} - p_{-} - p') 
\frac{d^3 p_{+}}{2 \omega_{+} (2 \pi)^3}
\frac{d^3 p_{-}}{2 \omega_{-} (2 \pi)^3}
\frac{d^3 p'}{2 \omega' (2 \pi)^3}  
 \times   \frac{1}{flux}
\overline{ | {\cal M}^{\gamma p \to M^+ M^- p} |^2} \;, 
\label{2_to_3}
\end{equation}
\end{widetext}
where for  photoproduction,
$flux = 4 \sqrt{(p \cdot q)^2 - m_p^2 q^2 } = 2 (s - m_p^2)$ and 
for a three-body reaction the amplitude $[ {\cal M} ] $ carries
dimension of $\mbox{ GeV}^{-1}$.  The unpolarized cross section is
a function of the following variables, 
the square of the center of mass energy, 
$s=(q+p)^2$, the  two-meson invariant mass, 
$M_{MM}$,  $M_{MM}^2 = (k_{+} + k_{-})^2$,
the four-momentum transfer squared in the nucleon,  $t = (p - p')^2$
and the polar  and azimuthal angles,
$\Omega = (\theta,\phi)$ specifying the direction of momentum of one
of the two produced mesons in the the two-meson recoil center of mass
(RCM) system. The coordinates of the RCM system are usually chosen
such that the $y$ axis is perpendicular to the production plane
determined by the photon and the recoil nucleon momenta, and
 $z$ is chosen either in the direction of the photon momentum
(in the RCM system), which is then  referred to as the
Gottfried-Jackson (GJ), or $z$ is  anti-parallel to
the recoiling nucleon direction, which defines the so called s-channel 
helicity (SCH).  At high energy ($s \gg M_{MM}^2, t$) the cross section takes a simple form
\begin{equation}
\frac{d \sigma ( M_{MM}^2,t,\theta, \phi)} {d M_{MM}^2 dt d \Omega} =
\frac{\beta}{16 \pi^4} \frac{1}{s^2}  \;
\overline{ | {\cal M}^{\gamma p \to M^+ M^- p} |^2}  \; ,
\label{master_formula}
\end{equation}
where $\beta = \sqrt{1 - \left(\frac{2 m_M}{M_{MM}}\right)^2}$
is the magnitude of velocity of each  meson in the center of mass of
the M$^+$M$^-$ system. The invariant amplitude for the
$2 \to 3$ continuum process can be written in the Regge factorized form 
 corresponding to the diagrams (a) and (b) in Fig.~\ref{fig_diagrams},  
%----------------------------

\begin{widetext}
\begin{equation}
{\cal M}_{\lambda_{\gamma} \lambda \to \lambda'}^{\gamma p \to M^+ M^- p}
(s,t,s_{+},t_{+},s_{-},t_{-}) =  
 = V_{\gamma M^+}(\lambda_{\gamma})
\frac{F_{os}(t_{+})}{t_{+}-m_{M}^2} 
{\cal M}_{\lambda \lambda'}^{M^- p}(s_{-},t) 
+
V_{\gamma M^-}(\lambda_{\gamma})
\frac{F_{os}(t_{-})}{t_{-}-m_{M}^2}
{\cal M}_{\lambda \lambda'}^{M^+ p}(s_{+},t). 
\label{Regge_factorization}
\end{equation}
\end{widetext}
In the equation above $s_{+}$ and $s_{-}$ are the Mandelstam variables
for the $M^+ p$ and $M^- p$ elastic scattering and
$t_{+}$ and $t_{-}$ are squares of momenta of the virtual
mesons ($t_{\pm} = (k_{\pm}-q)^2$). For pseudoscalar mesons, because
$q \epsilon_{\pm}( \lambda_{\gamma} = \pm 1 ) = 0$,
the corresponding vertex functions read
\begin{equation}
\begin{split}
V_{\gamma M^{\pm}}(\lambda_{\gamma}) = \pm e \; (2 k_{\pm}^{\mu}) \;
\epsilon_{\mu}(\lambda_{\gamma}) \; .
\end{split}
\label{vertices}
\end{equation}
The matrix elements take a simple form
in the Gottfried-Jackson frame, 
\begin{equation}
\begin{split}
\epsilon(\lambda_{\gamma} = \pm 1) \cdot k_{+} = \mp \frac{k}{\sqrt{2}}
\sin(\theta_{GJ}) \exp(\pm i \phi_{GJ} ) \; \\
\epsilon(\lambda_{\gamma} = \pm 1) \cdot k_{-} = \pm \frac{k}{\sqrt{2}}
\sin(\theta_{GJ}) \exp(\pm i \phi_{GJ} ) \; \\
\end{split}
\end{equation}
The denominators in Eq.(\ref{Regge_factorization}) can be calculated
in terms of the canonical variables $t$, $M_{MM}$ and cos$\theta$ as
\begin{equation}
t_{\pm} - m_M^2  = -2 q k_{\pm} = 
- \frac{1}{2} (M_{MM}^2 - t) (1 \mp \beta cos \theta) \; .
\end{equation}
For small invariant masses $M_{M^+ M^-}$ of the two-meson state,
(in the case of light meson production) sufficiently large overall $s$ 
 leads to large $s_{+}$ and $s_{-}$ in the relevant subprocesses.
At the high energies the invariant amplitudes for
the $2 \to 2$ quasi-elastic subprocesses can be written in the
simple but rather accurate form,
\begin{equation}
{\cal M}_{\lambda \lambda'}^{M^{\pm} p}(s_{\pm},t) = 
i s_{\pm} \sigma^{tot}_{M^{\pm} p}(s_{\pm})
\exp\left( \frac{B}{2} t \right)
\; \delta_{\lambda \lambda'} \; .
\label{2_to_2}
\end{equation}
The Kronecker $\delta_{\lambda \lambda'}$ reflects explicit
imposition of the target nucleon helicity conservation, known to hold
at high energies. The total cross section for $\pi p$ and $K p$ 
are well known \cite{PDG} and at high energies one can use
the Donnachie-Landshoff parametrizations \cite{DL92}.
For heavy mesons an educated guess will be necessary.
In general, the heavier the mesons, the smaller the corresponding
total cross section. The factor $F_{os}(t_{\pm})$ in Eq.(\ref{Regge_factorization}) takes into account the extended nature of the exchanged particle. Since for the process considered there are two vertices with an off-shell pseudoscalar meson, it is natural to write the combined form factor in the factorized form:
\begin{equation}
F_{os}(t_{\pm}) = F_{em}^{hos}(Q^2,t_{\pm},m_M) \cdot F_{corr}(t_{\pm}) \; ,
\end{equation}
where the first factor is the half-off-shell electromagnetic
form factor from the upper vertex and the second one is a form factor 
 representing the middle vertices in diagrams (a) and (b) in Fig.\ref{fig_diagrams}.
The exact form of the form factors is not known. In principle, a good quality data would help to find the proper functional form.

%The vertex function for real (transverse) photons is
%
%\begin{equation}
%V_{\gamma M^{\pm}} = ........ \; .
%\label{vertex}
%\end{equation}
Energy conservation imposes natural limits on energies
in different two-body subsystems
\begin{equation}
\begin{split}
W_{+} \equiv \sqrt{s_+} < W - m_M \; , \\
W_{-} \equiv \sqrt{s_-} < W - m_M \; , \\
M_{MM} < W - m_p \; .
\end{split}
\end{equation}
The amplitude given in Eq.(\ref{Regge_factorization}) is not yet
complete since it does not satisfy electromagnetic current
conservation. This current is given by, 
\begin{widetext}
\begin{equation}
J^\mu = 2 e \left[ k^\mu_+  
\frac{F_{os}(t_{+})}{t_{+}-m_{M}^2} 
{\cal M}_{\lambda \lambda'}^{M^- p}(s_{-},t) 
-  k^\mu_-  \frac{F_{os}(t_{-})}{t_{-}-m_{M}^2}
{\cal M}_{\lambda \lambda'}^{M^+ p}(s_{+},t) \right]
\label{jmu}
\end{equation}
\end{widetext}
so that the amplitude in Eq.(\ref{Regge_factorization})
can be written as, 
\begin{equation}
{\cal M}_{\lambda_{\gamma} \lambda \to \lambda'}
^{\gamma p \to M^+ M^- p}= 
 \epsilon_\mu(\lambda_\gamma) J^\mu(s,t,s_{+},t_{+},s_{-},t_{-}) 
\end{equation}
Current conservation implies $q_\mu J^\mu = 0$ while from
Eq.(\ref{jmu}) we find
\begin{eqnarray}
q_\mu J^\mu & = &  -4e \left[  
 F_{os}(t_{+}) {\cal M}_{\lambda \lambda'}^{M^- p}(s_{-},t)  \right. 
 \nonumber \\
& &  \left. - F_{os}(t_{-}){ \cal M}_{\lambda \lambda'}^{M^+ p}(s_{+},t) \right]. 
\end{eqnarray}
The origin of current non-conservation is two-fold. It comes form
the non-point-like nature of the exchanged particles which introduces
the from factors,  $F_{os} \ne 1$ and from the difference
in meson-nucleon scattering for the two charged mesons.
The later implies that electromagnetic charge flows differently
in $M^+p \to M^+p$ and $M^-p \to M^-p$ subprocesses and since photon
couples to all charge currents there has to be a correction which
reflects this difference. 
 
Having identified the two sources which contribute to the current we
can unambiguously find the required corrections.  We want to separate
the corrections to the current which arise from interactions in
the upper (meson) and lower (baryon) vertices and therefore we define, 
\begin{equation}
F_{\pm} \equiv {1\over 2} \left[ F_{os}(t_{+}) \pm  F_{os}(t_{-}) \right] 
\end{equation}
and 
\begin{equation}
M_{\pm} \equiv   = {1\over 2} \left[   {\cal M}_{\lambda \lambda'}^{M^- p}(s_{-},t)  \pm 
 {\cal M}_{\lambda \lambda'}^{M^+ p}(s_{+},t) \right].  
\end{equation}
The current can be written as
\begin{equation}
 J^\mu = J^\mu_C + J^\mu_N + J^\mu_M
\end{equation}
where
 \begin{equation}
 J^\mu_C =  2e\left[   {{k^\mu_+} \over { t_+-m_M^2}}  
 -   {{ k^\mu_-} \over {t_- - m_M^2}}  \right]
 ( F_+ M_+ + F_- M_-)
 \end{equation}
 is a conserved current, 
 \begin{equation}
 J^\mu_N = 2e \left[ {{k^\mu_+} \over { t_+-m_M^2}}  
 +   {{ k^\mu_-} \over {t_- - m_M^2}}  \right] F_+ M_- 
 \end{equation}
 is non-conserved due to a difference between $M^+N$ and $M^-N$
cross-sections, and
 \begin{equation}
 J^\mu_M =   2e \left[ {{k^\mu_+} \over { t_+-m_M^2}}  
 +   {{ k^\mu_-} \over {t_- - m_M^2}}  \right] F_- M_+ 
  \end{equation}
is non-conserved due the extended structure of the exchanged meson.  Now the additional contribution to the current required by current
conservation will depend on meson variables for $J^\mu_M$ and nucleon
variables for $J^\mu_N$, respectively, 
\begin{equation}
J^\mu_N \to J^\mu_N + \delta J^\mu_N 
\end{equation}
with
\begin{equation}
\delta J^\mu_N = 2 e {{ (p + p')^\mu } \over { q(p+p')}}  F_+ M_- 
\end{equation} 
and
\begin{equation}
J^\mu_M \to J^\mu_M + \delta J^\mu_N
\end{equation}
with
\begin{equation}
\delta J^\mu_N = 2 e {{ (k_+ + k_-)^\mu } \over {q(k_+ + k_-)}}  F_- M_+ 
=  -4 e {{ (k_+ + k_-)^\mu } \over { t - M_{MM}^2 } }  F_- M_+ \label{djn}
\end{equation}
The the unphysical pole at $t = M_{MM}^2$, Eq.(\ref{djn}) should be  eliminated by a zero in  the $F_-$ form factor vanishes at $t = M_{MM}^2$. 

\begin{figure}[htb] % Figure 2
\begin{center}
\includegraphics[width=7cm]{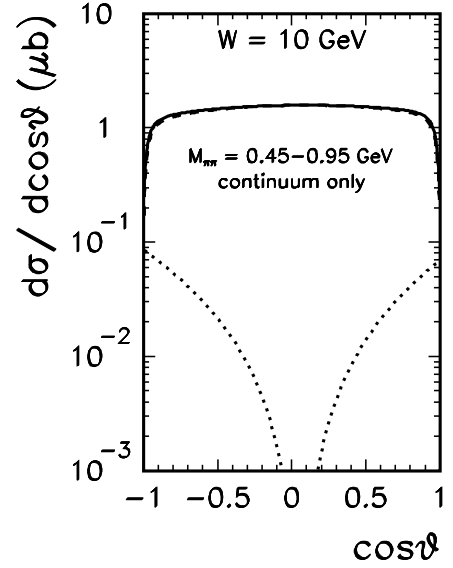}
\caption{
Angular distribution of $\pi^+$ in the Gottfried-Jackson frame.
The dashed line is the contribution without corrections restoring
current conservation. 
The dotted line represent the cross section associated with the
corrections itself and the solid line corresponds to the full,
current conserving amplitude. In this calculation we used the monopole
off-shell form factor with $M_{os}$ = 1 GeV
(see Eq.(\ref{monopole})).
\label{fig_cc}
}
\end{center}
\end{figure}

In Fig.\ref{fig_cc} we show the effect of these corrections on the angular distribution for $\pi^+\pi^-$ production calculated  in the GJ frame. 
The correction is generally very small, except for the tips of the angular distributions.

%-----------------------------------------------------------------------

%-----------------------------------------------------------------------

In this example we have integrated over the invariant mass range
 $M_{\pi\pi}$ = 0.45--0.95 GeV and over the kinematically accessible
range of the momentum transfer $t$. 
It is easy to check that the correction is negligible
 for the invariant mass distribution.

\begin{figure}[htb] % Figure 3
\begin{center}
\includegraphics[width=6cm]{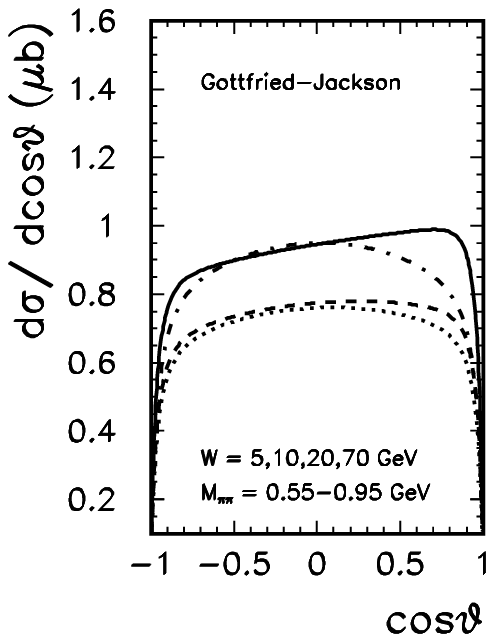}
\includegraphics[width=6cm]{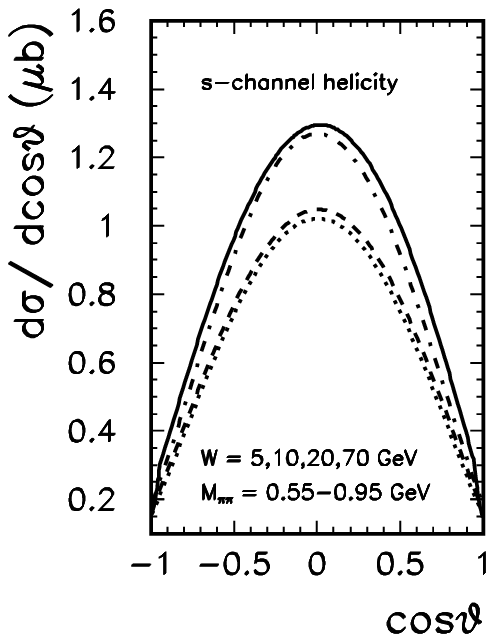}
\caption{
Angular distributions of $\pi^+$ from continuum amplitude 
in the Gottfried-Jackson (left panel) and in the s-channel helicity
(right panel) frames for four different energies:
W = 5 GeV (solid), W = 10 GeV (dashed),
W = 20 GeV (dotted) and W = 70 GeV (dash-dotted).
In this calculation we used the monopole
off-shell form factor with $M_{os}$ = 1 GeV
(see Eq.(\ref{monopole})).
\label{fig_gj_sch}
}
\end{center}
\end{figure}

In this paper we shall concentrate on the distributions
in $M_{MM}$ and cos$\theta$.
Often one is interested in distributions in the s-channel helicity
frame and not in the Gottfried-Jackson frame.
From the definition of the two frames it follows that the 
spherical angles in the GJ frame can be expressed by those in
the SCH frame 
${\cal M}_{\lambda_{\gamma}}^{\gamma p \to M^{+}M^{-}p}
\left(t,M_{MM},\theta_{GJ},\phi_{GJ} \right)$, 
with $\theta_{GJ} =\theta_{GJ}(\theta_{SCH},\phi_{SCH})$ and $\phi_{GJ} = \phi_{GJ}(\theta_{SCH},\phi_{SCH})$ through a rotation around the $y$ axis by an angle
$\theta_{rot}$~\cite{Sch}  which in the high energy limit is given by, 
\begin{equation}
\cos \theta_{rot} = \frac{M^2_{MM}+t}{M^2_{MM}-t} \; .
\end{equation}
In Fig.\ref{fig_gj_sch} we compare angular distributions
for the continuum in the Gottfried-Jackson (left panel) and
in the s-channel helicity (right panel) frames.
The shapes in both frames are rather different.
The effect of the rotation was neglected in the early
calculations.

%------------------------------------------------------------------

%------------------------------------------------------------------------

In order for the Regge parametrization of the two-body amplitudes
to be reliable, for $\pi^+\pi^-$ production the energies
$W_{+}, W_{-}$ have to be at least larger than 2 GeV. 
 This can be fulfilled for center of mass energy W $>$ 4 GeV, i.e.
$E_{\gamma} >$ 8 GeV. The future GlueX experiment at TJNAF
is therefore at the border of application of the present model.
%For example, the present experiments at TJNAF or the old experiment
%for kaon pair production \cite{DESY_KK,Daresbury_KK} have too small
%energies to apply the present formalism.

%----------------
\section{Results}
%----------------

The cross section for a three-body reaction depends on five independent
kinematical variables. For the reactions considered it is customary to
use $M_{MM}$, $t$, $\theta$ and $\phi$ and calculate
${d \sigma(M_{MM},t,\Omega)}/{d M_{MM} dt d \Omega}$.
The invariant mass distribution is then obtained by integrating over
the remaining variables
\begin{equation}
\frac{d \sigma}{d M_{MM}} =
\int_{t_{min}(M_{MM})}^{t_{max}(M_{MM})} dt d \Omega
\; \frac{d \sigma(M_{MM},t,\Omega)}{d M_{MM} dt d \Omega} \; ,
\label{invariant_mass_dist}
\end{equation}
where $t_{min}$ and $t_{max}$ are calculated from the three-body
kinematics (see for instance \cite{BK_book}).

%-----------------------------
\subsection{ Light meson pairs}
%-----------------------------

Below $M_{\pi\pi}$ = 1 GeV the $\rho^0$ meson dominates 
 the two-pion invariant mass spectrum.
The amplitude for the resonance is taken in the 
relativistic Breit-Wigner form.
%\footnote{ There is no complete agreement as to what is the best
%parametrization for broad resonance amplitude and 
% different parametrizations have been used in the literature.}
In our simple approach the normalization constant at $t=0$ is fixed
based on the vector meson dominance model. We write the resonant
three-body amplitude as,
\begin{widetext}
\begin{equation}
{\cal M}_{\lambda_{\gamma} \lambda \to \lambda'}^
{\gamma \to \rho^0 \to \pi^+\pi^-}
(s,t,M_{\pi\pi},\theta,\phi) =
C_{conv} \;
\frac{e}{\gamma_{\rho}} \;
{\cal M}_{\lambda \lambda'}^{\rho^0 p}(s,t) \; 
\; f_{BW}(M_{\pi\pi}) Y_{1,\lambda_{\gamma}}(\theta,\phi) \; .
\label{rho0_amplitude}
\end{equation}
\end{widetext}
Here we have introduced the amplitude for
quasi-elastic scattering of $\rho^0$ meson off the proton
\begin{equation}
{ \cal M}_{\lambda \lambda'}^{\rho^0 p}(s,t) =
 i s \sigma_{\rho^0 p}^{tot}(s) \exp\left( \frac{B_{\rho p} t}{2} \right) \; 
\delta_{\lambda \lambda'} \; .
\end{equation}
As for the continuum model the Kronecker $\delta_{\lambda \lambda'}$
reflects high-energy helicity conservation in the proton vertex.
The factor $C_{conv}$ is adjusted to reproduce the correct normalization 
of the amplitude.  

Different normalization of the reaction amplitudes are used
in the literature. Throughout the present paper we use a popular in
high-energy diffraction (see e.g. \cite{BP}) normalization,
such that the angular distribution for the two-body reaction is
\begin{equation}
\frac{d \sigma}{d \Omega_{CM}} = \frac{1}{64 \pi^2 s}
\left( \frac{p_f}{p_i} \right) \overline { | {\cal M}_{fi} |^2 }
\label{amplitude_normalization_1}
\end{equation}
and the optical theorem at high energy reads
\begin{equation}
Im {\cal M}(s,t=0) = s \sigma^{tot}(s)  \; .
\label{amplitude_normalization_2}
\end{equation}
This fixes the $C_{conv}$ factor in Eq.(\ref{rho0_amplitude}).

The factor $f_{BW}$ is
the standard relativistic Breit-Wigner propagator, 
\begin{equation}
f_{BW}(M_{\pi\pi}) = \frac{\sqrt{M_0 \Gamma(M_{\pi\pi})/\pi } }
{M_0^2 - M_{\pi \pi}^2 - i M_0 \Gamma(M_{\pi\pi}) }
\; ,
\label{f_BW}
\end{equation}
where $\Gamma(M_{\pi\pi}) = \Gamma_0
 \left(
\frac{M_{\pi\pi}^2-4 m_{\pi}^2}{M_0^2-4 m_{\pi}^2}
\right)^{3/2}$.
and it is normalized according to 
$\int |f_{BW}(M_{\pi \pi})|^2 \; dM_{\pi \pi}^2$ = 1 if $\Gamma(M_{\pi\pi}))$ is replaced by the 
energy independent width, $\Gamma_0$. 
We take $M_0$ = 768 MeV and $\Gamma_0$ = 151 MeV and 
\begin{equation}
\sigma_{\rho^{0}p}^{tot}(s) = \frac{1}{2}
 \left(
\sigma_{\pi^+p}^{tot}(s) + \sigma_{\pi^-p}^{tot}(s) 
 \right) \; .
\label{sigtot_rho0p}
\end{equation}

As for the continuum model the total cross sections
for $\pi^+p$ and $\pi^-p$  are taken from
the Donnachie-Landshoff parametrization \cite{DL92}.
In the present approach we assume the same
slope parameter for the resonance and continuum contributions,
$B_{\rho p} = B_{\pi p} \equiv B$.
%Because in our model both the continuum and resonance contributions
%are treated on the same footing we may expect correct
%proportions of the two processes.
Except for the off-shell dependence determined by the form factors
$F_{os}$ our model is essentially parameter free.
We used different parametrizations the form factor: 
 the exponential form,  
\begin{equation}
F_{os}(t_{\pm}) = \exp
\left( \frac{  t_{\pm} - m_{M}^2  } {2 \Lambda_{os}^2}  \right)
\label{exponential}
\end{equation}
and the monopole form
\begin{equation}
F_{os}(t_{\pm}) = \frac{M_{os}^2 - m_{M}^2}{M_{os}^2 - t_{\pm}} \; .
\label{monopole}
\end{equation}
The form factor is normalized to unity at the on-shell point
$t = m_M^2$. The exponential form is useful as the universal parameter $\Lambda_{os}$ for different meson exchanges can be used. In the monopole parametrization $M_{os} > m_M$, {\it i.e.}  different cut-off parameters for different exchanges have to be used. On the other hand the monopole form seems to be preferred, because at small virtualities
vector dominance applies and in addition it produces a correct
pQCD dependence at large virtualities.

\begin{figure}[htb] % Figure 4
  \subfigure[]{\label{fig_expo_old}
    \includegraphics[width=6.cm]{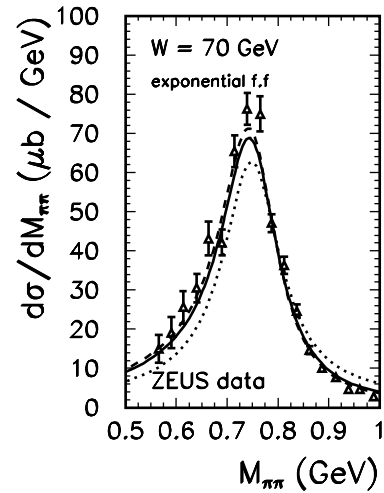}}
  \subfigure[]{\label{fig_mono_old}
    \includegraphics[width=6.cm]{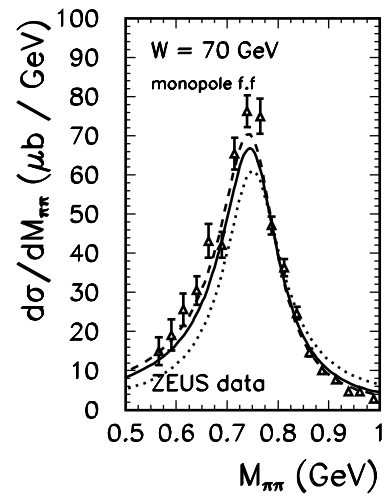}}
\caption{
The spectrum of invariant mass $M_{\pi \pi}$ for W = 70 GeV
for exponential and monopole off-shell form factor
without rotation of continuum
arguments.
The experimental results of the ZEUS collaboration are
from \cite{ZEUS_Mpipi}.
In this calculation B = 8 GeV$^{-2}$.
\label{fig_expo_vs_mono}
}
\end{figure}

\begin{figure}[htb] % Figure 5
  \subfigure[]{\label{fig_expo_new}
    \includegraphics[width=6.cm]{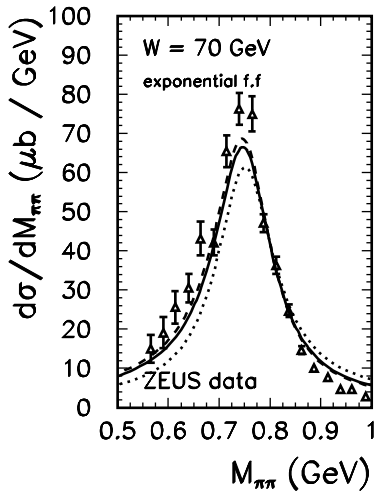}}
  \subfigure[]{\label{fig_mono_new}
    \includegraphics[width=6.cm]{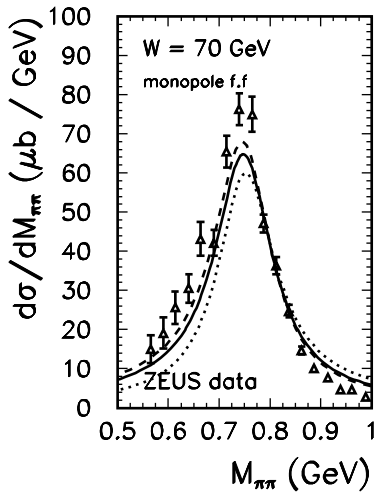}}
\caption{ Same as in Fig.~\ref{fig_expo_vs_mono} 
 but with rotation of continuum
arguments. 
\label{fig_expo_vs_mono2}
}
\end{figure}

The off-shell form factor is the least  know element of our model.
In Figs.\ref{fig_expo_vs_mono},~\ref{fig_expo_vs_mono2} we present spectral shape (resonance + continuum) for the two choices 
 of from factors 
 without and with rotation from SCH to GJ frame.  Somewhat better agreement is obtained without performing the
extra rotation of arguments. 

%------------------------------------------------------------------------

%------------------------------------------------------------------------

%------------------------------------------------------------------------

\begin{figure}[htb] % Figure 6
\begin{center}
\includegraphics[width=6cm]{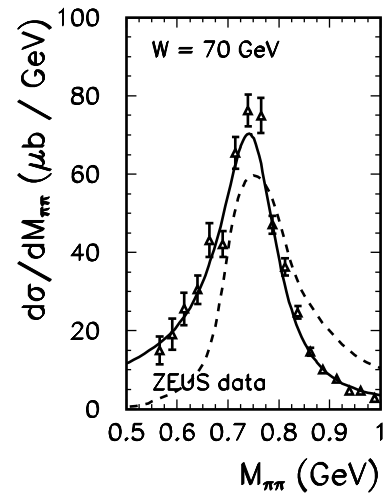}
\caption{
The spectrum of invariant mass $M_{\pi \pi}$ for W = 70 GeV.
The experimental results of the ZEUS collaboration are from \cite{ZEUS_Mpipi}.
In this calculation B = 8 GeV$^{-2}$. The standard resonance
contribution is shown as a  reference (dashed line).
\label{fig_m_pipi}
}
\end{center}
\end{figure}

%We do not find any simple explanation
%of this fact.
The spectral shape depends on the value of the form factor
parameter ($\Lambda_{os}$ = 0.5, 1.0, 2.0 GeV for the exponential form factor, and 
$M_{os}$ = 0.5, 1.0, 2.0 GeV for the monopole form factor).
For pion production very similar results are obtained with
exponential and monopole form factors for $\Lambda_{os} \approx M_{os}$.
Therefore, having in view a possible universality of the exponential
off-shell form factor parameter $\Lambda_{os}$, we shall use
 the exponential form factor in the following.
The experimental data from the ZEUS collaboration at DESY
\cite{ZEUS_Mpipi} are superimposed on the theoretical lines.
In the absence of other mechanisms the value of the off-shell
form factor could be obtained from the fit to the experimental
data.
The coherent sum of the resonance and continuum (solid line) differs
considerably from the standard resonance shape which is
shown in Fig.\ref{fig_m_pipi}. The resonance contribution
alone (dashed line) gives a poor description of the data.
In particular, the position of the maximum is at higher $m_{\pi\pi}$
then observed experimentally. These features have been often 
 ignored in the literature and only integrated
cross section were used to compare with theoretical calculations.

%------------------------------------------------------------------

%------------------------------------------------------------------------

%The same is true with the shift of the peak position
%compared to other experiments on $\rho$ production.

\begin{figure}[htb] % Figure 7
\begin{center}
\includegraphics[width=6cm]{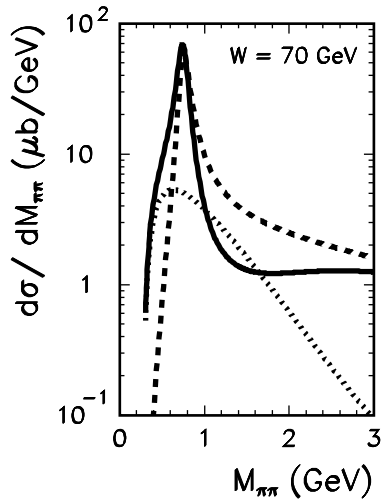}
\caption{
The decomposition of the spectrum of the $\pi^+\pi^-$ invariant mass
$M_{\pi \pi}$ for W = 70 GeV. We show separately the
resonance (dashed line) and continuum (dotted line)
cross sections. The solid line corresponds to
a coherent sum of both processes processes. 
In this calculation B = 8 GeV$^{-2}$.
\label{fig_decompo}
}
\end{center}
\end{figure}

In order to have a better insight into the origins of the line-shape modifications  in Fig.\ref{fig_decompo}
we show separately the resonance and the continuum contributions
in somewhat broader range of two-pion invariant mass. This figure
clearly demonstrates that it is mainly the interference effect which
deforms the spectral shape of the $\rho$ meson. 

%------------------------------------------------------------------

%------------------------------------------------------------------

In our approach
the background is described by a physical process which has 
a strong two-pion invariant mass dependence. Often  
in experimental analyses \cite{ZEUS_Mpipi, ZEUS_t, ZEUS_Q2}
(see also \cite{RS98}), this background is parametrized with
a weak $\pi\pi$ mass dependence. 
We do not expect that our model could describe the experimental
spectra above $M_{\pi\pi} >$ 1 GeV, since there will be important
contributions from higher mass two-pion resonances
{\it e.g.} $f_2(1270)$, $\rho_3(1690)$, $\rho(1700)$. 
The situation there may be therefore rather complicated and
we leave the corresponding analysis for future investigations. 
%Can our continuum be a dominant effect for $M_{\pi\pi} >$ 2 GeV?
%Also this problem goes beyond the scope of our present paper
%as it requires a knowledge of tails of broad resonances below.

\begin{figure}[htb] % Figure 8
\begin{center}
\includegraphics[width=6cm]{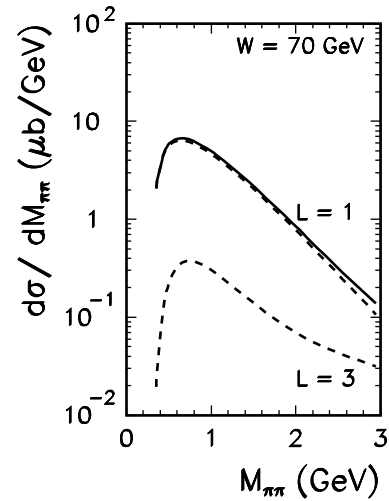}
\caption{
The contributions of individual partial waves to the spectrum
of invariant mass $M_{\pi \pi}$ for W = 70 GeV (excluding resonance
production). 
In this calculation B = 8 GeV$^{-2}$.
\label{fig_l_deco}
}
\end{center}
\end{figure}

In order to further understand the large interference effect of
the continuum contribution with the $P$-wave resonance
we performed a decomposition of the continuum amplitude into
the partial wave series in the GJ frame, 

\begin{eqnarray}
{\cal M}^{\gamma p \to M^+ M^- p}_{\lambda_{\gamma},\lambda \to \lambda'}
(t, & & \!\!\!\!\!\!M_{\pi \pi},\theta,\phi) = \nonumber \\
& & =\sum_{l,m}
a_{lm}^{\lambda_{\gamma},\lambda,\lambda'}(t,M_{\pi\pi})
\;
Y_{lm}(\theta,\phi). \nonumber \\
\label{partial_wave_deco}
\end{eqnarray}
The expansion coefficients can be calculated as
\begin{eqnarray}
& & a_{lm}^{\lambda_{\gamma},\lambda,\lambda'}(t,
M_{\pi\pi}) = \nonumber \\
& & = \int Y_{lm}^*(\theta,\phi) \cdot
{\cal M}^{\gamma p \to M^+ M^- p}_{\lambda_{\gamma},\lambda \to \lambda'}
(t,M_{\pi \pi},\theta,\phi) \;
d\Omega . \nonumber \\
\label{alm}
\end{eqnarray}
In our model the expansion coefficients depend only on
$\lambda_{\gamma},l,m$, i.e.
\begin{equation}
a_{lm}^{\lambda_{\gamma},\lambda,\lambda'}(t,M_{\pi\pi}) \equiv
a_{lm}^{\lambda_{\gamma}}(t,M_{\pi\pi}) \;.
\label{dummy_indices}
\end{equation}
We find that in the case of the pseudoscalar production
the continuum contributes dominantly to the $P$-wave, 
 {\it i.e.} 
$|a_{1m}| \gg |a_{00}|, |a_{2m}|$, etc.  and 
$|a_{1-1}^{+1}| < |a_{10}^{+1}| < |a_{1+1}^{+1}|$,and $|a_{1-1}^{-1}| > |a_{10}^{-1}| > |a_{1+1}^{-1}|$. 
This explains the large interference between the continuum
and resonance production.  We also find a relatively large
contribution of the $F$-wave.
The individual contributions of $l=1$ and $l=3$ partial wave are shown
in Fig.\ref{fig_l_deco}.

%------------------------------------------------------------------------

%------------------------------------------------------------------------

%In the searches for new states in a two-meson subsystem
%of a three- or four-body final state, the $F$-wave contributions
%are often neglected in the partial wave analysis (PWA).
% Their inclusion could would significantly complicate the PWA.
Presence of the $F$-wave is interesting
in the context of forward-backward asymmetry and the  moment analysis 
(see e.g.\cite{BLS04}). These are usually discussed
in terms of the $S$- and $P$- wave interferences. Our analysis shows
that in principle one needs to include $S, P, D$ and $F$ waves
into such an analysis even for relatively low invariant masses.

Let us try to understand this hierarchy of the partial wave amplitudes. 
 The angular distribution originates from 
\begin{eqnarray}
{\cal M}^{\gamma p \to M^+ M^- p}_{\lambda_{\gamma}} (t,& & \!\!\!\!M_{MM},\theta,\phi) \nonumber  \\
& & \propto q_{MM} i {\cal F}(t) \exp(\pm i \phi) \sin \theta 
{\cal A}(\theta,\phi), \nonumber \\
\label{Deck_schematic}
\end{eqnarray}
where we have defined a slowly changing function of $\theta$
and $\phi$,
\begin{eqnarray}
{\cal A}(\theta,\phi) &  = &  
\frac{\sigma_{M^-p}(s_{-}(\theta,\phi)) F_{os}(t_{+}(\theta))}
{1-\beta \cos \theta} \nonumber \\ 
& + & 
\frac{\sigma_{M^+p}(s_{+}(\theta,\phi)) F_{os}(t_{-}(\theta))}
{1+\beta \cos \theta}.
\label{slow_function}
\end{eqnarray}
It then becomes obvious that it is the function ${\cal A}(\theta,\phi)$
which is responsible for generation of partial waves different than
$l = 1$. It is easy to show that if the numerators were identical the
$S$-wave amplitude would vanish. 
The smooth energy dependence of the cross sections introduces
a small $S$-wave contribution. This small effect is slightly dependent on
the incident energy.

%One general comment is in order here.
The large interference effect is specific to the photoproduction
of $P$-wave resonances and the $P$-wave dominated pseudoscalar meson continuum. 
%We do not expect a strong interference effect between, for example, %$S$-wave or $D$-wave resonances in photoproduction.
%The continuum photoproduction of other (nonpseudoscalar mesons)
%will have different partial wave decomposition and may
%lead to strong interference effects also for resonances
%with $l \ne$ 1.

\begin{figure}[htb] % Figure 9
\begin{center}
\includegraphics[width=6cm]{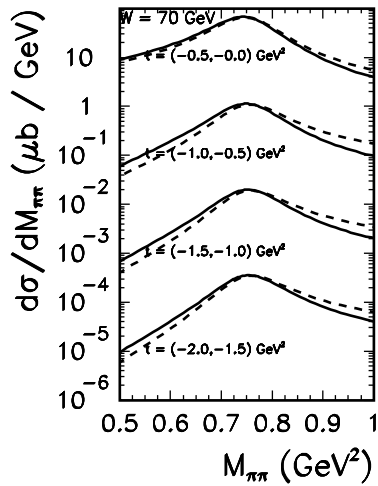}
\caption{
The spectrum of invariant mass $M_{\pi \pi}$ for W = 70 GeV
for  different bins in $t$. The solid lines correspond to the results
without and the dashed line to the results with the rotation of
arguments of the continuum contribution.
In this calculation B = 8 GeV$^{-2}$ and exponential off-shell form
factor with $\Lambda$ = 1 GeV has been used.
\label{fig_t_pipi}
}
\end{center}
\end{figure}

In Fig.\ref{fig_t_pipi} we discuss the evolution of the spectral
shape (asymmetry) as a function of the momentum transfer, $|t|$.
While at low $|t|$ the spectral asymmetry is reversed compared
to the standard resonant shape at large $|t|$ one observes a restoration
of the standard asymmetry expected for the $P$-wave resonance.
Such an effect was observed experimentally in \cite{ZEUS_t} and
to the best of our knowledge the dynamics of this effect
has not been given before. 

%------------------------------------------------------------------------

%------------------------------------------------------------------------

So far we have neglected the final state
$\pi \pi$ interaction. This can be restored by modifying the continuum partial-wave amplitudes, 
\begin{equation}
\tilde{a}^{\lambda_\gamma,\lambda,\lambda'}_{lm}
= a^{\lambda_\gamma,\lambda,\lambda'}_{lm}
\cos \delta_l^{\pi \pi}
\exp( i \delta_l^{\pi \pi} ) \; .
\label{pipi_fsi}
\end{equation}
Here $\delta_l^{\pi \pi}$ is the
$M_{\pi \pi}$-dependent phase shift for $\pi \pi$ scattering in 
the $l$'th partial wave. If only resonant final state interaction effects are included the modified partial amplitude of the continuum becomes, 
\begin{equation}
\tilde{a}^{\lambda_\gamma,\lambda,\lambda'}_{lm}
= a^{\lambda_\gamma,\lambda,\lambda'}_{lm}
\left(
\frac{ M_{\pi \pi}^2 - M_0^2 }{ M_{\pi \pi}^2 - M_0^2 + i M_0 \Gamma }
\right) \; .
\label{resonant_fsi}
\end{equation}
Such a modified amplitude vanishes at $M_{\pi \pi} = M_0$.
For $l=1$ the final amplitude can be written as a sum of three terms:
the direct $\rho^0$ production, free meson pair production and
the resonance  production via re-scattering. The last two are
contained in Eq.(\ref{resonant_fsi})  (or Eq.(\ref{pipi_fsi})).  
We checked that  re-scattering from the continuum back to the 
resonance  is negligible.

\begin{figure}[htb] % Figure 10
\begin{center}
\includegraphics[width=6cm]{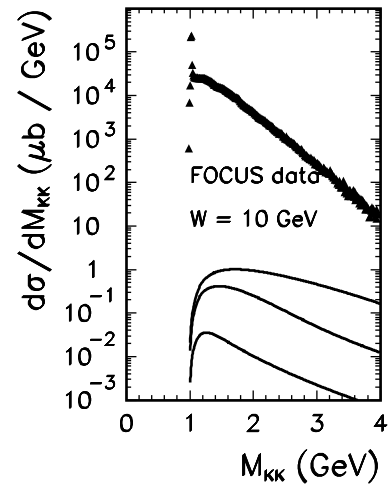}
\caption{
The spectrum of invariant mass $M_{K{\bar K}}$. The experimental results
of the FOCUS collaboration (without acceptance corrections)
are from \cite{Mitchell_PhD}. The experimental
data is not normalized.
In this calculations $B$ = 6 GeV$^{-2}$ and
and exponential off-shell form factor with $\Lambda$ = 0.5, 1.0, 2.0
GeV has been used.
\label{fig_m_KK}
}
\end{center}
\end{figure}

The $K^+ K^-$ production is more sensitive to the choice
of the off-shell form factor.
The spectrum of two-kaon invariant mass is shown
in Fig.\ref{fig_m_KK} for exponential form factor and 
$\Lambda_{os}$ = 0.5, 1.0, 2.0 GeV.

%------------------------------------------------------------------------

Recent experimental data from the FOCUS
collaboration are shown for comparison \cite{FOCUS_KK}.
Our model of diffractive production of kaonic pairs gives
a good description of the main trend of the data.
A direct comparison with the FOCUS data is, however, not possible,
because the experimental acceptance is known only in very
limited range of the phase space.
The absolutely normalized experimental invariant mass distribution
would help to better limit the only unknown parameter
$\Lambda_{os}$ (or $M_{os}$) of
the off-shell form factor. The situation may be, however, not so simple
in light of resent experimental data for $\gamma \gamma$
collisions \cite{BELLE_KK}.
In principle, similarly as for the $\pi^+ \pi^-$ invariant mass
the interference of the $K^+ K^-$ continuum and the $\phi$ resonance 
 takes place. However, because the $\phi$ has a much smaller
decay width than the $\rho^0$ and is situated very close
to the $K^+ K^-$ threshold (where the continuum contribution is small),
its importance for $M_{KK} >$ 1.1 GeV is negligible.

%The future GlueX experiment 
%allow to study angular distributions of the $\pi^+ \pi^-$
%and $K^+ K^-$ photoproduction. 
The angular distributions
depend on the invariant mass. Thus one studies angular
distributions in bins of invariant masses.
The angular distribution is calculated from the 4-dimensional
differential cross section as,
\begin{equation}
\frac{d \sigma} {d cos\theta} = 
\int_{M_{min}}^{M_{max}} d M_{MM}
\int_{t_{min}}^{t_{max}} dt \int d\phi
\; \frac{d \sigma(M_{MM},t,\Omega)}{d M_{MM} dt d \Omega} \; .
\label{angular}
\end{equation}

\begin{figure}[htb] % Figure 11
\begin{center}
\includegraphics[width=6cm]{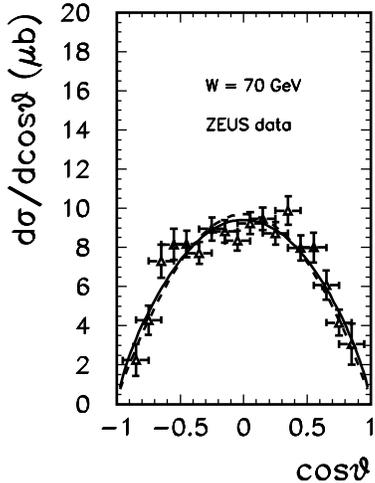}
\caption{
The angular distribution of charge pions in the SCH frame 
 for W = 70 GeV and $M_{\pi\pi}$ in the measured range.
The experimental results of the ZEUS collaboration are
from \cite{ZEUS_Mpipi}.
In this calculation B = 8 GeV$^{-2}$ and exponential off-shell form
factor with $\Lambda$ = 1 GeV have been used.
The solid line corresponds to the calculation without rotation
of arguments of the continuum, whereas the dashed line corresponds
to the calculation with the rotation.
\label{fig_zeus_cos}
}
\end{center}
\end{figure}

In Fig.\ref{fig_zeus_cos} we present angular distribution
for the ZEUS kinematics \cite{ZEUS_Mpipi}. The angular distribution
is almost proportional to $\left( \sin\theta \right)^2$. The continuum
contribution only slightly modifies the $\left( \sin\theta \right)^2$
resonance distribution. The results with or without rotation between the frames are almost indistinguishable.  

%------------------------------------------------------------------

%------------------------------------------------------------------------

\begin{figure}[htb] % Figure 12
\begin{center}
\includegraphics[width=6cm]{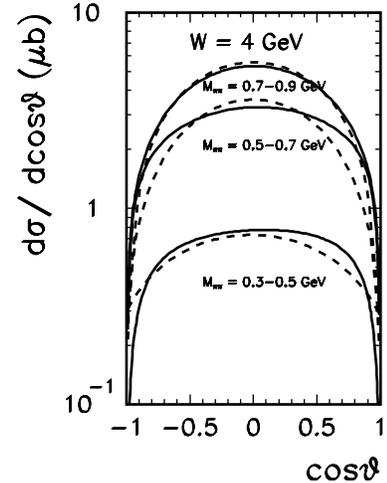}
\caption{
The angular distribution of $\pi^+$ in the SCH frame for 
the GlueX energy W = 4 GeV. In this calculation $B$ = 6 GeV$^{-2}$
and exponential off-shell form factor with $\Lambda$ = 1 GeV
have been used. The sold line corresponds to the case of no extra
rotation while dashed line corresponds to the result with
the extra rotation.
\label{fig_zpipi04}
}
\end{center}
\end{figure}

In Fig.\ref{fig_zpipi04} we present predictions for
angular distributions for the GlueX experiment at TJNAF.
The shape of the distributions depends on the interval of
the two-pion invariant mass. The effect of the rotation
of the arguments of the continuum is stronger for smaller
invariant masses. We predict a sizable asymmetry with, $\pi^+$ being preferentially emitted in the forward (photon) direction.

%------------------------------------------------------------------------

%---------------------------------------------------------------------

\begin{figure}[htb] % Figure 13
\begin{center}
  \subfigure[]{\label{fig_afb_pi_a}
    \includegraphics[width=6.0cm]{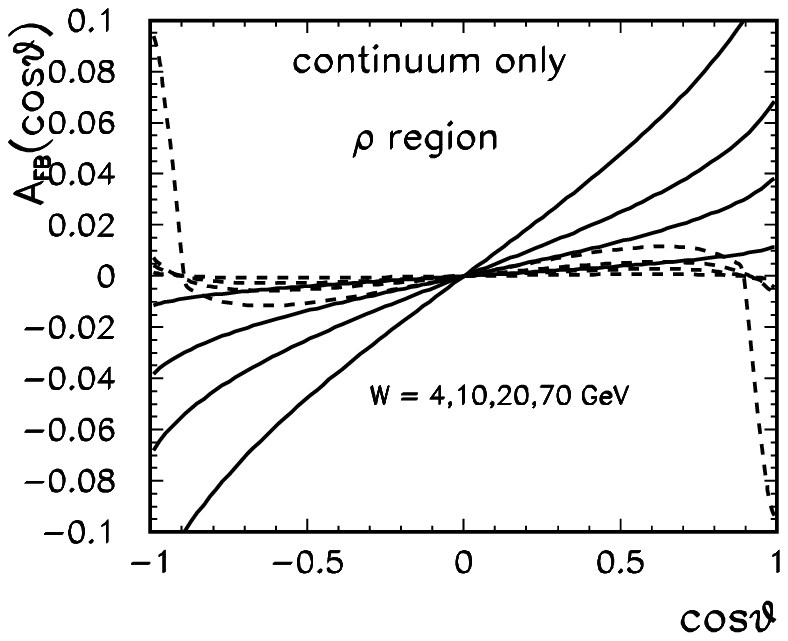}}
  \subfigure[]{\label{fig_afb_pi_b}
    \includegraphics[width=6.0cm]{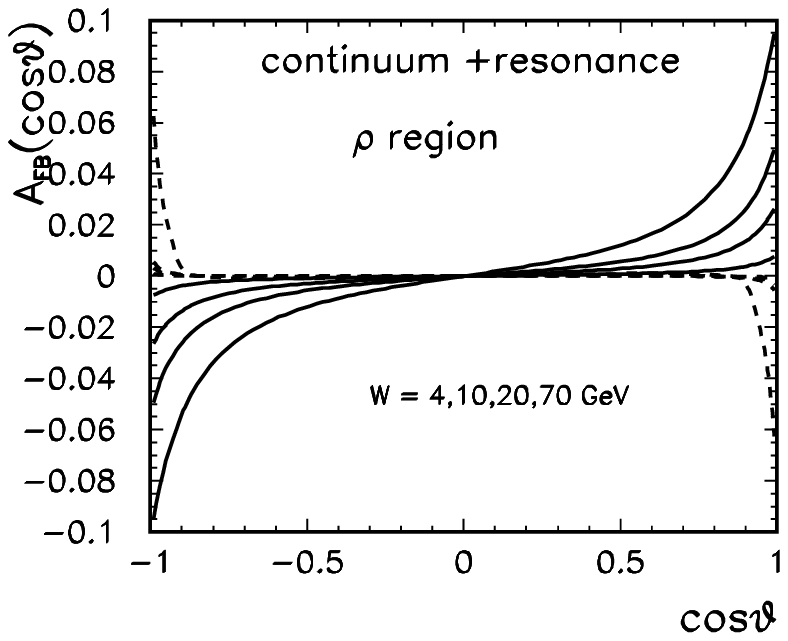}}
\caption{
Forward-backward asymmetry for $\pi^+$ for the continuum
(panel a) and for the resonance+continuum (panel b)
as the function of cos$\theta$ for four different energies:
W = 4, 10, 20, 70 GeV.
The solid line corresponds to the continuum in the GJ, while
the dashed line to the continuum in the SCH frame.
The kinematical cuts are specified in the text.
\label{fig_afb_rho}
}
\end{center}
\end{figure}

The function ${\cal A}(\theta,\phi)$ in Eq.(\ref{Deck_schematic})
is responsible not only for generating higher partial waves
but also for the forward-backward asymmetry.
In order to measure the asymmetry we define the following quantity, 
\begin{equation}
A_{FB}^{\pi^{\pm}}(\theta) \equiv
\frac
{  \frac{d \sigma^{\pi^{\pm}}}{dz} \left( \theta \right)
 - \frac{d \sigma^{\pi^{\pm}}}{dz} \left( \theta \right) }
{  \frac{d \sigma^{\pi^{\pm}}}{dz} \left( \theta \right)
 + \frac{d \sigma^{\pi^{\pm}}}{dz} \left( \theta \right) }
\; .
\label{fb_asymmetry}
\end{equation}
By construction
\begin{equation}
\frac{d\sigma}{d \theta}(\theta_{\pi^+}) =
\frac{d\sigma}{d \theta}(\pi - \theta_{\pi^-}) \; ,
\label{symmetry_of_ad}
\end{equation}
which means that the asymmetry must fulfil the symmetry relations
\begin{equation}
\begin{split}
A_{FB}^{\pi^+}(\theta) = - A_{FB}^{\pi^-}(\theta) \; , \\
A_{FB}^{\pi^{\pm}}(\theta) = - A_{FB}^{\pi^{\pm}}(\pi - \theta) \; .
\end{split}
\end{equation}
The asymmetry for $\pi^+$ is shown in Fig.\ref{fig_afb_rho}
for four different energies and for one bin in $M_{\pi\pi}$, 
 0.55 GeV $ < M_{\pi\pi} <$ 0.95 GeV and $|t| <$ 1 GeV$^2$. 
  We present separately the asymmetry of the continuum contribution alone (panel a) and the asymmetry
of the sum of the resonance and continuum contributions (panel b).
A sizable asymmetry can be seen in the GJ frame (solid lines).
In general, the larger incident energy, the smaller the asymmetry.
The inclusion of the resonance contribution lowers the asymmetry
around $z$ = 0. 

%---------------------------------------------------------------------

%---------------------------------------------------------------------

For the  GlueX experiment at TJNAF
we predict the asymmetry to be about 10\% at the very forward and very backward directions. Even at $W = 70 \mbox{ GeV}$ 
 the asymmetry is of the order of 1\%.  In our model the asymmetry is caused
by the different interaction of $\pi^+ p$ and $\pi^- p$
(analogously for $K^+ p$ and $K^- p$). This is caused
by a different strength of sub-leading reggeons. 
The asymmetry disappears at large energy where the dynamics
of elastic scattering is governed exclusively by the pomeron exchange.
The asymmetry discussed above is an important
test of the diffractive mechanism. 
%To the best of our knowledge it was not
%thoroughly discussed in the literature before.
When transformed to the SCH frame the asymmetry becomes rather
negligible (dashed lines).

In Refs.\cite{HPST02,GIN03} the observation of forward-backward
asymmetry of charged pions was proposed in order to pin down
 the  odderon exchange.
These analyses were based on the assumption that only resonant
mechanisms plays a role. The mechanism considered here
was not taken into account. Our  diffractive mechanism may mimic
the pomeron-odderon interference effects discussed in
Refs.\cite{HPST02,GIN03}. A careful search for the odderon
exchange must therefore necessarily include the two-pion continuum
discussed in the present paper.

\begin{figure}[htb] % Figure 14
\begin{center}
\includegraphics[width=6cm]{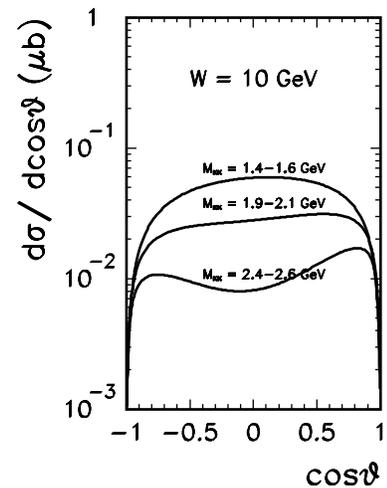}
\caption{
The angular distribution of $K^+$ in the GJ frame for a typical
FOCUS energy W = 10 GeV. In this calculation $B$ = 6 GeV$^{-2}$. 
\label{fig_zkk10}
}
\end{center}
\end{figure}

We expect that the diffractive production of $K^+ K^-$
is the dominant mechanism well above the $\phi$ resonance.
In Fig.\ref{fig_zkk10} we present angular distributions
of $K^+$ in the GJ recoil center of mass system for a typical
FOCUS energy, $W=10 \mbox{ GeV}$ for different bins of $M_{KK}$ specified in the figure. 

%------------------------------------------------------------------

%---------------------------------------------------------------------

As for the pion production
we obtain asymmetric distributions with $K^+$ being
preferentially emitted in the forward hemisphere.
The shape of the angular distribution changes with $M_{KK}$.
The shape is very important when studying resonances.
For example, the FOCUS collaboration  found difficulties
 in spin assignment of the  X(1750) \cite{Mitchell_PhD}.
This may partially be due to the interference of the resonance
 and the continuum. 
In general, the larger the invariant mass, the more the cross
section peaks at forward/backward directions.
This is related to the increase in the  number of active partial waves
with the increasing subsystem energy.

\begin{figure}[htb] % Figure 15
\begin{center}
\includegraphics[width=6cm]{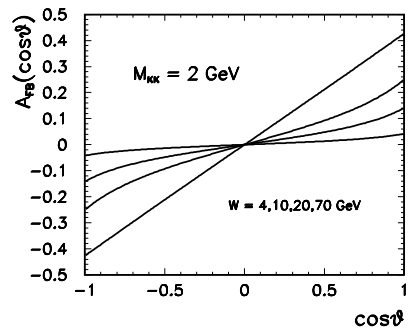}
\caption{
Forward-backward asymmetry for $K^+$ production in the GJ frame
as the function of cos$\theta$ for four different
energies: W = 4, 10, 20, 70 GeV.
The kinematical cuts are specified in the text.
\label{fig_afb_kk}
}
\end{center}
\end{figure}

In Fig.\ref{fig_afb_kk} we present $A_{FB}$ for $K^{+}$
for four different incident energies, 1.9 GeV $< M_{KK} <$ 2.1 GeV
and $|t| <$ 1 GeV$^2$. We observe much larger asymmetries
than for the $\pi^+ \pi^-$ case. This is partially due to larger
asymmetries in $K^+ p$ and $K^- p$ scattering than
for the $\pi^+ p$ and $\pi^- p$ scattering. 

%------------------------------------------------------------------

%---------------------------------------------------------------------

%The second
 %reason is that the  $\rho^0$ resonance contribution which by itself is %symmetric.
In principle the $K^+ K^-$ asymmetry should be seen
in the data of the FOCUS collaboration. The possible higher mass resonance contributions are expected to lower the asymmetry.
A careful search for the asymmetry as a function of the two-kaon
invariant mass would be very interesting in searches
for new states. 

%-----------------------------
\subsection{Heavy meson pairs}
%-----------------------------

\begin{figure}[htb] % Figure 16
\begin{center}
\includegraphics[width=6cm]{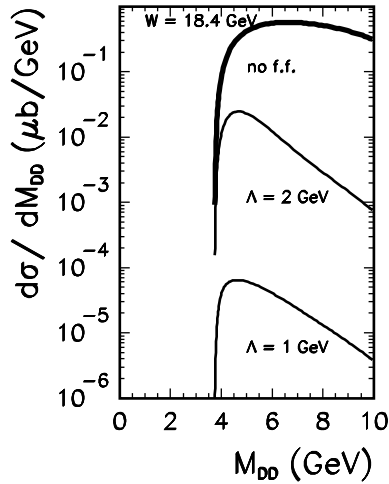}
\caption{
The spectrum of invariant mass $M_{DD}$. The theoretical results
were obtained with B = 6 GeV$^{-2}$ and exponential
off-shell form factor with parameters specified in the figure.
\label{fig_m_DD}
}
\end{center}
\end{figure}

In Fig.\ref{fig_m_DD} we show invariant mass distributions of
diffractively produced pairs of $D^+ D^-$ for the average energy
of the FOCUS experiment \cite{FOCUS_DD}.

%---------------------------------------------------------------------

%---------------------------------------------------------------------

The averaging is not so important as the diffractive contribiution
is only slowly dependent on the incident energy.
We show results for different values of the scale parameter
 in the exponential off-shell form factor.
The absolute normalization strongly depends on the value of
the form factor mass parameter. This is due to the fact that
the intermediate meson D in Fig.\ref{fig_diagrams} is far from its
mass shell.
Our distributions peak at $M_{DD}$ = 4.5--5 GeV.
In principle, the FOCUS collaboration could analyze their data
and try to construct the invariant mass distribution.
The expected statistics will be of course rather low, of the order of
10-50 events.

\begin{figure}[htb] % Figure 17
\begin{center}
\includegraphics[width=6cm]{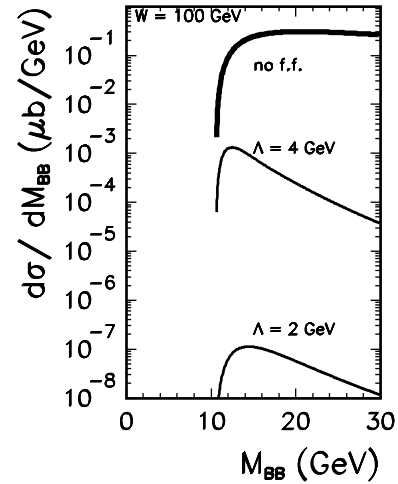}
\caption{
The spectrum of invariant mass $M_{BB}$. The theoretical results
were obtained with B = 6 GeV$^{-2}$ and exponential
off-shell form factor with parameters specified in the figure.
\label{fig_m_BB}
}
\end{center}
\end{figure}

The invariant mass distribution for the $B^+ B^-$ pair production is
shown in Fig.\ref{fig_m_BB} for a typical HERA energy $W = 100 \mbox{ GeV}$. Here there is a dramatic effect on the value of the form factor parameter. The distribution reaches its maximum at rather high $M_{BB}$. The smallness of the cross sections precludes, however, studies
of differential cross section. 

%---------------------------------------------------------------------

%---------------------------------------------------------------------

\begin{figure}[htb] % Figure 18
\includegraphics[width=6cm]{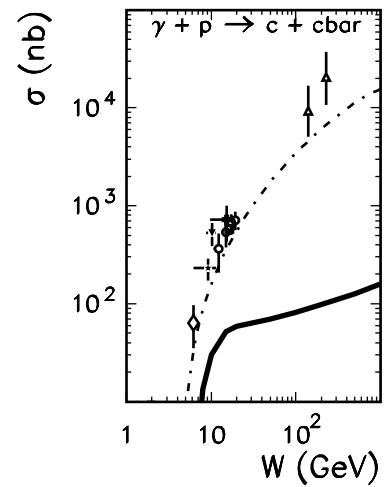}
%  \subfigure[]{\label{fig_w_ccbar_low}
 %   \includegraphics[width=7.0cm]{gp_ccbar.eps}}
 % \subfigure[]{\label{fig_w_bbbar}
 %   \includegraphics[width=7.0cm]{gp_bbbar.eps}}
\caption{
The energy dependence of the integrated cross section for
the diffractive photoproduction of
$D^+ D^-$ (thick solid line) with our estimate
of the upper limit ($\Lambda_{os}$ = 2 GeV).
For comparison we show the cross section
for open charm and bottom production (dash-dotted line) and
corresponding experimental data.
The details concerning the open heavy-flavour production
can be found in Ref.\cite{Szczurek_QQbar}. 
\label{fig_heavy_W}
}
\end{figure}

\begin{figure}[htb] % Figure 19
\includegraphics[width=6cm]{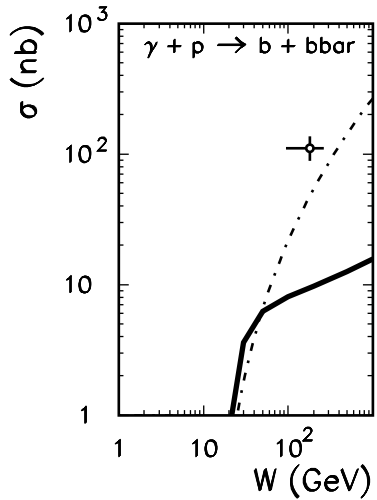}  %\subfigure[]{\label{fig_w_ccbar_low}
 %   \includegraphics[width=7.0cm]{gp_ccbar.eps}}
 % \subfigure[]{\label{fig_w_bbbar}
 %   \includegraphics[width=7.0cm]{gp_bbbar.eps}}
\caption{ Same as in Fig.~\ref{fig_heavy_W} for $B^+B^-$ production,  ($\Lambda_{os}$ = 4 GeV).   
\label{fig_heavy_W1} }
\end{figure}

In Figs.\ref{fig_heavy_W},~\ref{fig_heavy_W1} we show the energy dependence of the integrated
cross section for both $DD$ and $BB$ pair production (thick solid line).
For comparison we show the standard collinear-factorization results
for open charm and open bottom production \cite{Szczurek_QQbar}
as well as the experimental cross sections. As can be seen from the figure
the diffractive cross section is much smaller than its counterpart
for the standard photon-gluon fusion followed by fragmentation.

%---------------------------------------------------------------------

%---------------------------------------------------------------------
%---------------------------------------------------------------------

%---------------------------------------------------------------------

In addition, the rise of the cross section with energy is
slower then for the photon-gluon fusion.
Only close to the kinematical threshold the diffractive component
may constitute a sizable fraction of the open heavy-flavor component.
Since in this case application of the meson-exchange approach is
less reliable  than for the light pairs, an experimental measurement
would be very interesting and helpful to discriminate between models. 
  While for the $D^+ D^-$ pairs
 this may be feasible, it will be hard to expect such an analysis for the $B^+ B^-$ pairs,
at least in a few year perspective. We predict the fraction of
diffractive-to-nondiffractive events to be somewhat larger for bottom than for charm mesons.

%--------------------
\section{Conclusions}
%--------------------

We have presented a 
model of diffractive photoproduction of opposite-charge
pseudoscalar mesons. The model should be valid at sufficiently
high energies, down to energies relevant at the future
experiments in Hall D at Jefferson Lab. The understanding
of the two-meson continua is absolutely crucial when
looking for new (exotic) mesonic states in the two-meson channels.
Our model results seem to be consistent with the current experimental data. 
%A few examples were given. We have also presented also
% few interesting predictions.

The interference of the two-pion continuum and the resonant
$\rho^0$ contribution leads to a significant deformation
of the resonance peak, which is consistent with the experimental
data from the ZEUS collaboration at HERA \cite{ZEUS_Mpipi}.
The effect of the continuum-resonance interference is often
completely neglected. We explain a change of the spectral
shape of the $\rho^0$ bump with momentum transfer. 

%We get a good description of the slope of the experimental distribution
%measured some time ago by the OMEGA collaboration at CERN \cite{Omega}. 

The diffractive mechanism of the $K^+ K^-$ production leads
to a broad bump in invariant mass distribution, above the $\phi$
meson position with the maximum at $M_{KK} \approx$ 1.4 GeV.
Experimentally, except for the $X(1750)$ mentioned earlier no clear resonances in the $K^+ K^-$ channel have been seen in \cite{FOCUS_KK}
{\it e.g.} even though  states like $a_2$, $f_2$, $\rho_3$ are expected. 
A careful phase-shift analysis will be required to separate them. 
Our model amplitude provides a reasonably well controlled
background for such studies.

We have found a forward-backward asymmetry in $\pi^+$ and $\pi^-$
(similarly $K^+$ and $K^-$) recoil center of mass angular
distributions.
The new effect is due to different interaction of
different-charge mesons with the proton. The effect disappears at large
incident energies, where the pomeron exchange is dominant. 
%We hope that this effect can be identified
% experimentally even at present by the FOCUS collaboration.
This asymmetry may be very important in the context of recently
proposed searches for odderon exchange via forward-backward
charge asymmetry in the region of $M_{\pi\pi} \sim$ 1 -- 2 GeV.

The cross sections for diffractive production of heavy meson
pairs were neither calculated nor measured in the past.
In the present paper we have presented
an estimates of the corresponding cross sections.
The cross sections obtained in our analysis are, however, very small. 
Even a measurement of  integrated cross section for diffractive
$D^+ D^-$ pair photoproduction would be very useful for
testing the mechanism of diffractive production.
Such data would allow one to pin down off-shell
effects for the intermediate (exchanged) heavy meson.
%We hope that the FOCUS collaboration at Fermilab could make
%an event-by-event analysis of their present data
%in order to extract the discussed by us diffractive cross section
%for $D^+ D^-$ production.

\begin{acknowledgments}
We thank Ryan Mitchell for providing us the experimental
two-kaon invariant mass distribution of the FOCUS collaboration.
We are also indebted to Leonard Le\'sniak  and Geoffrey Fox for several discussions.
This work was partially supported by the US Department of Energy grant
DE-FG0287ER40365 (APS), and the Nuclear Theory Center at
Indiana University. 
\end{acknowledgments}
%------------------------------------------------------------------


\begin{thebibliography}{99}


\bibitem{Drell60}
S.D. Drell, Phys. Rev. Lett. {\bf 5}, 278 (1960).

%\bibitem{Deck64}
%R. Deck, Phys. Rev. Lett. {\bf 13} (1964) 169.

\bibitem{Soding65}
P. S\"oding, Phys. Lett. {\bf 19},  702 (1965).

\bibitem{Krass67}
A.S. Krass, Phys. Rev. {\bf 159},  1496 (1967).

\bibitem{KU69}
G. Kramer and J.L. Uretsky, Phys. Rev. {\bf 181},  181  (1969).

\bibitem{Pumplin70}
J. Pumplin, Phys. Rev. D{\bf 2}, 1859  (1970) .

\bibitem{KQ71}
G. Kramer and H.R. Quinn, Nucl. Phys. B{\bf 27}, 77  (1971).

\bibitem{PDG}
Review of Particle Physics, Particle Data Group,
Eur. Phys. J. C{\bf 15}, 1 (2000).

\bibitem{ZEUS_Mpipi}
M. Derrick et al. (ZEUS collaboration), Z. Phys.  C{\bf 69}, 39  (1995).

\bibitem{ZEUS_Q2}
J. Breitweg et al. (ZEUS collaboration), Eur. Phys. J.  C{\bf 6}, 603  (1999).

\bibitem{ZEUS_t}
J. Breitweg et al. (ZEUS collaboration), Eur. Phys. J.  C{\bf 14}, 213  (2000).

\bibitem{E852}
D.R. Thompson et al. (E852 collaboration), Phys. Rev. Lett.
{\bf 79}, 1630  (1997).

\bibitem{CB}
A. Abele et al. (Crystal Barrel collaboration), Phys. Lett. 
 B{\bf 423}, 175  (1998).

\bibitem{AS1}  A.~R.~Dzierba {\it et al.}, Phys.\ Rev.\ D {\bf 67},
094015 (2003). 

\bibitem{AS2}  A.~P.~Szczepaniak, M.~Swat, A.~R.~Dzierba and S.~Teige, Phys.\ Rev.\ Lett.\  {\bf 91}, 092002 (2003).

\bibitem{FOCUS_KK}
J.M. Link et al. (FOCUS collaboration), 
 Phys. Lett. B{\bf 545}, 50  (2002).

\bibitem{COMPASS}
A. Sandacz (COMPASS collaboration), private communication.

\bibitem{BELLE_KK}
K. Abe et al. (BELLE collaboration), Eur. Phys. J.  C{\bf 32}, 323  (2003).

\bibitem{JKLSW1998}
Ch-R. Ji, R. Kami\'nski, L. Le\'sniak, A. Szczepaniak
and R. Williams, Phys. Rev.  C{\bf 58}, 1205  (1998).

\bibitem{BLS04}
{\L}. Bibrzycki, L. Le\'sniak and A.P. Szczepaniak,
Eur. Phys. J.  C{\bf 34}, 335  (2004).

\bibitem{gp_bbbar}
S. Chekanov et al. (ZEUS collaboration), hep-ex/0405069.

\bibitem{gg_bbbar}
A. Csilling (OPAL collaboration), {\it in PHOTON2004,
Ambleside, UK, August 26-31, 2000, hep-ex/0011022};\\
M. Acciari et al. (L3 collaboration), Phys. Lett. B{\bf 503}, 10  (2001);\\
F. Kapusta (DELPHI collaboration), {\it in MESON2004, Cracow, Poland, June 4-8, 2004}.

\bibitem{FOCUS_DD}
J.M. Link et al. (FOCUS collaboration), Phys. Lett.  B{\bf 566}, 51  (2003).

\bibitem{LS04}
M. {\L}uszczak and A. Szczurek, 
Phys. Lett.  B{\bf 594}, 291  (2004).

\bibitem{BP}
V. Barone and E. Predazzi, {\ it High-Energy Particle Diffraction},
(Springer, Berlin 2002).

\bibitem{DL92}
A. Donnachie and P.V. Landshoff, Phys. Lett.  B{\bf 296}, 227  (1992).

\bibitem{Sch} K.~Schilling, P.~Seyboth and G.~E.~Wolf,
Nucl.\ Phys.\ B {\bf 15}, 397 (1970)
[Erratum-ibid.\ B {\bf 18}, 332 (1970)].

\bibitem{RS98} M.G. Ryskin and Yu.M. Shabelski, Phys. Atom. Nucl. {\bf 61}, 81  (1998).

\bibitem{DESY_KK}
D.C. Fries et al., Nucl. Phys.  B{\bf 143}, 408 (1978).

\bibitem{Daresbury_KK}
D.P. Barber et al., Z. Phys.  D{\bf 12}, 1 (1982).

\bibitem{BK_book}
E. Byckling and K. Kajantie, {\it Particle Kinematics},
(John Wiley and Sons, London 1973).

%\bibitem{Omega}
%D. Aston et al. (OMEGA collaboration), Phys. Lett. {\bf B92} (1980) 215.

\bibitem{Mitchell_PhD}
R.E. Mitchell, {\it Phd thesis, the University of Tennessee, Knoxville,
December 2003}.

\bibitem{HPST02}
Ph. Hagler, B. Pire, L. Szymanowski and O.V. Teryaev,
Eur. Phys. J.  C{\bf 26}, 261  (2002).

\bibitem{GIN03}
I.F. Ginzburg, I.P. Ivanov and N.N. Nikolaev,
Eur. Phys. J.  C{\bf 30}, 2  (2003).

\bibitem{Szczurek_QQbar}
A. Szczurek, Eur. Phys. J.  C{\bf 26}, 183  (2003).

\end{thebibliography}
\end{document}